\documentclass{amsart}
\usepackage{amssymb}

\def\proclaim #1. #2\par{\medbreak
\noindent{\bf#1.\enspace}{\sl#2}\par\medbreak}
\begin{document}

\date{January 29, 2004}
\author[Edward G. Effros]{Edward G. Effros\\\vspace{.2in}Dedicated to Richard
V. Kadison and Masamichi Takesaki\\ for transmitting von Neumann's vision
\\\vspace{.2in}Expanded Lecture Notes \\Frontiers in Mathematics \\Texas
A\&M, November, 2003 }
\thanks{The author was patially supported by an NSF grant}
\begin{abstract}
This is an informal guide to the history of Heisenberg's matrix
mechanics. It is designed for mathematicians with only a minimal
background in either physics or geometry, and it is based upon
Heisenberg's original arguments.
\end{abstract}

\title[Matrix Revolutions]{Matrix Revolutions\\
The Origin of Quantum Variables}
\maketitle


\section{Introduction}

The most dramatic shift in Twentieth Century physics stemmed from
Heisenberg's formulation of matrix mechanics \cite{H1}. In classical
physics, quantities such as position, momentum, and energy are regarded as 
\emph{\ functions}. In quantum theory one replaces the functions by
non-commuting \emph{infinite matrices }or to be more precise, self-adjoint
operators on Hilbert spaces. This enigmatic step remains the most daunting
obstacle for those who wish to understand the subject.

Although there exist many excellent mathematical introductions to quantum
mechanics (see, e.g., \cite{M}, \cite{V}) they are understandably focused on
the development of mathematically coherent methods. As a result, mathematics
students must postpone understanding \emph{why} non-commuting variables
appeared in the first place. To remedy this, one can adopt a more historical
approach, such as that found in Emch's beautiful historical monograph \cite
{E}, the entertaining yet informative ``comic book'' \cite{L}, or Born's
classic text \cite{B}.

In recounting the creation of quantum mechanics, the most difficult task is
to describe how Heisenberg found the canonical commutation relation 
\begin{equation}
PQ-QP=\frac{h}{2\pi i}I  \label{comm}
\end{equation}
for the position and momentum operators $Q$ and $P$. This equation is the
final refinement of Planck's principle that a certain action variable is
discrete, or more precisely that it can assume only integer increments of a
universal constant $h$. Heisenberg used the more sophisticated formulation
of Bohr and Sommerfield that for periodic systems one has the ``quantum
condition'' 
\begin{equation}
\oint pdq=nh  \label{sommer}
\end{equation}
(see \S4).

In the words of Emch (see \cite{E}, p. 262) ``one can only propose some very
loose a priori justifications'' for the derivation of (\ref{comm}) from (\ref
{sommer}). Even Born, who was apparently the first to postulate the general form of
(\ref{comm}) (see \cite{E}, p. 264), avoided discussing it, appealing
instead to the Schr\"{o}dinger model (\cite{B}, p.130, see also \cite{L}
p.224), and this is the approach that one finds in most physics texts. We
will attempt to make Heisenberg's direct conceptual leap a little less
mysterious, by deciphering an argument that Heisenberg presented in his 1930
survey \cite{H2}. At the heart of his computation is the observation that 
\begin{quote}
\emph{the analogue of the derivative for the discrete action
variable is just the corresponding finite difference quotient.} 
\end{quote}
(see (\ref{difference})). 

Shortly after Heisenberg introduced matrix mechanics, Schr\"{o}dinger found
an alternative quantum theory based upon the study of certain wave equations 
\cite{S}. His approach enabled one to avoid a direct reference to
Heisenberg's matrices. Although it is both intuitive and computationally
powerful, ``wave mechancs'' is not as useful in quantum field theory. The
difficulty is that it does not fully accomodate the particle aspects of
quanta. In quantum field theory one must take into account the incessant
creation and annihilation of particles associated with the relativistic
equivalence of mass and energy. In particular, the number of particles
present must itself be regarded as an integer valued quantum variable. In
Born's words \cite{B}, p. 130, ``Heisenberg's method turns out to be more
fundamental.''

Our goal has been to maximize the accessiblity of the material. In order to
do this we have taken liberties with the mathematical, physical, and
historical details. To some extent this is justified by the fact that
regardless of how much care we might take, the discussion is necessarily
tentative. Although Heisenberg's argument is mathematically quite
suggestive, in the end we must discard these notions in favor of the
operator techniques that grew out of them.

\section{Atomic spectra, Fourier series and matrices}

The crisis that occured in classical physics is clearly seen in the peculiar
properties of atomic spectra. If one sends an electric discharge through an
elemental gas $A$ such as hydrogen or sodium, the gas will emit light
composed of very precise (angular) frequencies $\omega $. The corresponding 
\emph{spectrum }$\mathrm{sp}A$ of such frequencies is quite specific to the
element $A$. For a single frequency we have the corresponding representation 
\[
\cos (\omega t+a)=\mathrm{Re}\,e^{i(\omega t+a)}=c_{-1}e^{-i\omega
t}+c_{0}+c_{1}e^{i\omega t} 
\]
for suitable complex constants $c_{-1},$ $c_{0},$ $c_{1}.$ Superimposing
these frequencies, we may describe the radiation by the sum 
\begin{equation}
f_{A}(t)=\sum_{\omega \in \mathrm{sp}_{0}A}c_{\omega }e^{i\omega t},
\label{atoms}
\end{equation}
where $\mathrm{sp}_{0}A=\mathrm{sp}A\cup -\mathrm{sp}A\cup \{0\}.$

There are obvious classical analogues of this phenomenon. If one strikes an
object, the resulting sound can be decomposed into certain specific angular
frequencies. In the case of a tuning fork, the resulting motion is harmonic,
and one obtains a corresponding Fourier series for the amplitude of the
sound wave in the form 
\[
f(t)=A\cos (\omega t+a)=c_{-1}e^{-i\omega t}+c_{0}+c_{1}%
e^{i\omega t}. 
\]
for suitable complex coefficients $c_{k}$. If one instead plucks a guitar
string, the resulting sound is a combination of various frequencies, all of
which are overtones, i.e., multiples of a fundamental frequency $\omega .$
Thus one has a Fourier series 
\begin{equation}
f(t)=\sum_{n\in \Bbb{Z}}c_{n}e^{i(n\omega )t}.  \label{Fourier}
\end{equation}
where for simplicity we assume that only finitely many of the $c_{n}$ are
non-zero. We define the (full) \emph{spectrum} of $f$ to be the cyclic group 
$\Bbb{Z}\omega =\left\{ n\omega :n\in \Bbb{Z}\right\} $. As is well-known,
one can duplicate the sound of a guitar string by superimposing the pure
frequencies as in (\ref{Fourier}).

More complicated systems (such as a bell) will have several degrees of
freedom, and thus several fundamental frequencies. For a system with two
degrees of freedom one will have two fundamental frequencies $\omega ,\omega
^{\prime }$ with the ``almost periodic'' expansions

\[
f(t)=\sum_{n,n^{\prime }\in \Bbb{Z}}c_{n,n^{\prime }}e^{i(n\omega +n^{\prime
}\omega ^{\prime })t}. 
\]
We will restrict our attention to one degree of freedom.

The linear space 
$
\mathcal{A}(\omega )
$
of all functions of the form (\ref{Fourier}) with finitely many non-zero
terms is closed under multiplication since if we are given 
\begin{eqnarray*}
f(t) &=&\sum_{n\in \Bbb{Z}}c_{n}e^{i(n\omega )t} \\
g(t) &=&\sum_{n\in \Bbb{Z}}d_{n}e^{i(n\omega )t}
\end{eqnarray*}
then 
\[
f(t)g(t)=\sum_{k,n\in \Bbb{Z}}c_{k}d_{n-k}e^{i(k\omega )+((n-k)\omega
)t}=\sum_{n\in \Bbb{Z}}a_{n}e^{i(n\omega )t},
\]
where $a_{n}$ is the ``convolution'' 
\begin{equation}
a_{n}=(c*d)_{n}=\sum_{k\in \Bbb{Z}}c_{k}d_{n-k},  \label{convolution}
\end{equation}
Furthermore $\mathcal{A}(\omega )$ is closed under conjugation since 
\[
\bar{f}(t)=\sum c_{n}^{*}e^{i(n\omega )t}
\]
where $c_{n}^{*}=\overline{c_{-n}}.$ In more technical terms, the $*$
-algebra $\mathcal{A}(\omega )$ is a representation of the group $^{*}$%
-algebra $\Bbb{C[\Bbb{Z}}].$ This result, of course, stems from the fact
that $\mathrm{sp}f=\Bbb{Z\omega }$ is a group under addition.

Returning to atomic spectra, it is tempting to regard (\ref{atoms}) as some
kind of Fourier series. There are several problems with this interpretation.

First of all, we are actually interested in analyzing the property of a 
\emph{single atom}. In this case it is inappropriate to actually ``add up''
the series (\ref{atoms}). For example (getting a little ahead of ourselves)
a hydrogen atom will radiate only one frequency at a time corresponding to
the electron taking a particular orbital jump. Thus superpositions do not
occur when one ``watches'' a single atom. For this reason it is more
accurate to let $f(t)$ stand for the \emph{array} $(c_{\omega }e^{i\omega
t})_{\omega \in \mathrm{sp}_{0}A}$.

Secondly, in striking contrast to the classical models, it is not useful to
consider the additive group generated by $\mathrm{sp}_{0}A$. Given $\omega \in 
\mathrm{sp}_{0}A,$ one cannot expect to find \emph{any} of the overtones $%
n\omega $ in $\mathrm{sp}_{0}A.$ Nevertheless the set $\mathrm{sp}_{0}A$ does
display an exquisitely precise algebraic structure, called the \emph{\ Ritz
combination principle}. We may doubly index $\mathrm{sp}A$, i.e., we may let 
$\mathrm{sp}_{0}A$ $=\left\{ \omega _{m,n}\right\} _{m,n\in \Bbb{N}}$, in such a
manner that 
\begin{equation}
\omega _{m,n}+\omega _{n,p}=\omega _{m,p}  \label{Ritz}
\end{equation}
for all $m,n,p\in \Bbb{N}.$ In particular, $\omega _{m,m}+\omega
_{m,m}=\omega _{m,m}$ and thus $\omega _{m,m}=0.$ Furthermore, $\omega
_{m,n}+\omega _{n,m}=\omega _{m,m}=0$, and therefore $\omega _{n,m}=-\omega
_{m,n}$. Using this double indexing of the spectrum, our array becomes a 
\emph{matrix}: 
\begin{equation}
f(t)=[a_{m,n}e^{i\omega _{m,n}t}]_{m,n\in \Bbb{N}}.  \label{matrix}
\end{equation}

The set $\mathcal{M}(\omega )$ of matrices (\ref{matrix}) is already a
linear space. Owing to (\ref{Ritz}), $\mathcal{M}(\omega )$ is closed under
matrix multiplication and the adjoint operation since 
\begin{eqnarray*}
f(t)g(t) &=&\left[ \sum_{k}c_{m,k}e^{i\omega _{m,k}t}d_{k,n}e^{i\omega
_{k,n}t}\right] \\
&=&\left[ \sum_{k}c_{m,k}d_{k,n}e^{i(\omega _{m,k}+\omega _{k,n})t}\right] \\
&=&\left[ \sum a_{m,n}e^{i\omega _{m,n}t}\right]
\end{eqnarray*}
where $a=cd$ is the usual matrix product, and 
\[
f(t)^{*}=\left[ \bar{a}_{n,m}e^{-i\omega _{n,m}t}\right] =\left[
a_{m,n}^{*}e^{i\omega _{m,n}t}\right] 
\]
with $a^{*}$ the adjoint matrix. In fact one can regard $\mathcal{M}(\omega
) $ as a representation of the $*$-algebra $\Bbb{C}[\Bbb{N}\times \Bbb{N]}$
of the full groupoid $\Bbb{N}\times \Bbb{N}.$ This point of view has been
explored by Connes \cite{C}, but we will not pursue it further in this paper.

It is easy to prove that any doubly indexed family $\omega _{m,n}$
satisfying (\ref{Ritz}) must have the form 
\[
\omega _{m,n}=C_{m}-C_{n} 
\]
for suitable constants $C_{m}.$ The precise values for the hydrogen atom are
given by Balmer's equation 
\begin{equation}
\omega _{m,n}=2\pi R\frac{c}{m^{2}}-2\pi R\frac{c}{n^{2}}  \label{Balmer}
\end{equation}
where $c$ is the speed of light, and $R$ is known as the Rydberg's constant.

Long before matrices were introduced, Bohr justified Rydberg's equation by
combining Rutherford's model of the atom with a quantum condition on the
action variable. This ``old'' quantum theory was to play a crucial role in
the evolution of matrix mechanics.

\section{Action and quantization conditions}

Action is perhaps the least intuitive of the standard notions of classical
mechanics. As usual, the easiest way to understand a physical quantity is to
consider its units or ``dimensions''. We let $\mathcal{M},$ $\mathcal{L},$
and $\mathcal{T}$ denote units of mass $m$, length (or position) $q$, and
time $t$ (e.g. one can use grams, meters, and seconds). Given a physical
quantity $P$, we let $[P]$ denote its units. We have, for example, 
\begin{eqnarray*}
\lbrack \mathrm{velocity\,}v] &=&\left[ \frac{dq}{dt}\right] =\mathcal{L%
\mathcal{T}}^{-1} \\
\lbrack \mathrm{acceleration\,}a] &=&\left[ \frac{d^{2}q}{dt^{2}}\right] =%
\mathcal{L\mathcal{T}}^{-2} \\
\lbrack \mathrm{momentum\,}p] &=&\left[ mv\right] =\mathcal{ML\mathcal{%
\mathcal{T}}}^{-1} \\
\lbrack \mathrm{force\,}F\mathrm{\,}] &=&[ma]=\mathcal{MLT}^{-2} \\
\lbrack \mathrm{potential\,energy\,}V] &=&\left[ -Fq\right] =\mathrm{%
\mathcal{M}\mathcal{L}}^{2}\mathrm{\mathcal{T}}^{-2} \\
\left[ \mathrm{kinetic\,energy\,}T\right] &=&\left[ \frac{1}{2}mv^{2}\right]
=\mathrm{\mathcal{M}\mathcal{L}}^{2}\mathrm{\mathcal{T}}^{-2} \\
\left[ \mathrm{total\,energy\,}H\right] &=&[V+T]=\mathrm{\mathcal{M}\mathcal{%
L}}^{2}\mathrm{\mathcal{T}}^{-2}
\end{eqnarray*}
Noting that they have the same dimensions, we simply regard $V,$ $T,$ and $%
E=V+T$ as ``different forms'' of energy. We will often consider derivative
and integral versions of these quantities, such as $v$ and $a$ above and the
potential energy 
\[
V=-\int F(q)dq. 
\]
The dimensions frequently mirror physical laws. For example the equation for
force corresponds to Newton's second law. On the other hand the relativistic
equation $E=mc^{2}$ corresponds to $\mathrm{\mathcal{M\mathcal{L}}}^{2}%
\mathrm{\mathcal{T}}^{-2}=\mathcal{M}\times (\mathcal{L}\mathcal{T}%
^{-1})^{2} $.

The usual form of a travelling wave (in one spatial dimension) is given by 
\begin{equation}
f(t,q)=A\cos (\omega t+kq)  \label{wave}
\end{equation}
where $\omega $ is the angular frequency (radians per second) and $k$ is
angular wavenumber (radians per meter). The corresponding dimensions are

\begin{eqnarray*}
\lbrack \mathrm{angular\,frequency\,\,}\omega ] &=&[\mathrm{radians}]/\left[ 
\text{\textrm{time}}\right] =\mathcal{T}^{-1}. \\
\lbrack \mathrm{angular\,wavenumber\,\,}k] &=&[\mathrm{radians}]/[\mathrm{%
distance}]=\mathcal{L}^{-1}
\end{eqnarray*}
We recall these are related to the frequency $\nu $ (cycles per second) and
wavelength (of a cycle) $\lambda $ by $\omega =2\pi \nu $ and $k=2\pi
/\lambda l.$

Given an angular co-ordinate $\theta $ measured in radians, we have the
dimension 
\[
\lbrack \mathrm{angular\,\,velocity\,\,}\omega ]=\left[ \frac{d\theta }{dt}
\right] =\mathcal{T}^{-1} 
\]
By analogy with the momentum formula $p=mv,$ the angular momentum is defined
by $L=\iota\omega ,$ where $\iota$ is the ``moment of inertia'', or
equivalently $L$ is the signed length of the vector $\mathbf{L}=\mathbf{r}%
\times \mathbf{p,}$ where $\mathbf{r}$ is the position vector and $\mathbf{p}
$ is the momentum vector. Thus we have 
\[
\lbrack \mathrm{angular\,\,momentum}\,\,L]=\mathcal{ML}^{2}\mathcal{T}^{-1}. 
\]

In classical physics, the (restricted) action along a parametrized curve $%
\gamma $ is defined by the formulas 
\[
J[\gamma ]=\int_{\gamma }p\,dq=\int_{a}^{b}T\,dt, 
\]
and the actual motion taken by the particle is determined by finding the
stationary values of suitable variations of $J$ with fixed energy
(alternatively one can use a different variational principle involving the
Lagrangian, see \cite{Gel}, \cite{G}). The corresponding dimensions are
given by 
\[
\lbrack \mathrm{action\,}J]=[\mathrm{energy}]\times \left[ \text{\textrm{time%
}}\right] =[\mathrm{momentum}]\times \left[ \text{\textrm{distance}}\right] =%
\mathcal{ML}^{2}\mathcal{\mathcal{\mathcal{T}}}^{-1}. 
\]
We see from above that action has the same dimensions as angular momentum.
Following \cite{PR}, we will also use the action $I=(1/2\pi )J.$

Quantum mechanics began in 1900 with Planck's paper \cite{P}. He discovered
that he could predict the radiation properties of black bodies provided he
assumed a ``quantum condition''. He essentially postulated that the action
variable $J$ associated with an atom can take only the discrete values $%
nh, $ where $h$ is a universal constant and $n\in \Bbb{N}.$

An early task of quantum mechanics was to reconcile the particle and wave
properties of ``quantum objects'' such as photons and electrons. Einstein 
\cite{Ei} related the energy $E$ and momentum $p$ of a photon to the
frequency $\nu $ and the wavelength $\lambda $ of the corresponding wave.
Noting that $E/\nu $ and $p\lambda $ are action variables (see above), he
predicted that each of these equals the ``minimal action'' $h$, i. e., we
have the Einstein relations 
\begin{eqnarray*}
E &=&h\nu =\hbar \omega \\
p &=&h/\lambda =\hbar k
\end{eqnarray*}
where $\hbar =h/2\pi .$ Fifteen years later de Broglie \cite{D} proposed
that these relations were valid for all particles exhibiting the
wave-particle dichotomy, including the electron. It was a short step from
there to finding a wave equation for which the corresponding functions (\ref
{wave}) are solutions. This is precisely the Schr\"{o}dinger equation.

In 1913 Bohr used the Planck-Einstein quantum condition to explain the
spectral lines of the hydrogen atom \cite{Br}. He proposed that the electron
is constrained to particular circular orbits by the quantum condition. To be
more specific, he assumed that the electron had a specific energy $E_{m}$ in
the $m$-th orbit, and that it drops down (respectively jumps up) to the $n$%
-th orbit, it loses (respectively absorbs) energy $E_{m}-E_{n},$ which is
carried away or brought by a photon with frequency 
\begin{equation}
\omega _{m,n}=\frac{E_{m}-E_{n}}{\hbar }.  \label{Bohr}
\end{equation}
When Bohr used the classical Coulomb law to calculate the angular momentum $%
L $ of the electron in the $m$-th orbit, he discovered that it was given by $%
L=m\hbar $ for an integer $m.$ In fact by using the Hamiltonian theory from
the next section, he and Sommerfield showed that this coincides with
Planck's quantum condition $J=mh,$ and the latter is also true for arbitrary
closed orbital motions. Within a few years, Bohr's theory was used to
predict the frequencies of the spectral lines for a variety of systems.

Bohr also formulated a fundamental asymptotic property for the spectral
values which he called the \emph{correspondence principle}. Returning to the
Rydberg formula, he observed that for large $m,$ the electrons behaved
almost classically, in the sense that one obtained overtones. More
precisely, a drop of $k=m-n\ll m$ orbits resulted in the $k$-th overtone of
a fundamental frequency $\omega _{m}=2\pi Rc/m^{3}:$%
\begin{eqnarray*}
\omega _{m,m-k} &=&2\pi Rc(-\frac{1}{m^{2}}+\frac{1}{(m-k)^{2}}) \\
&=&2\pi k\left( \frac{Rc}{m^{3}}\right) \frac{(1-k /2m)}{(1-2k/m+k
^{2}/m^{2})} \\
&\sim &k\omega _{m}.
\end{eqnarray*}
A similar principle applies if $k$ is negative. We will use the notation $%
k\ll n$ to indicate relatively small positive or negative jumps.

In principle it would seem that we might have to consider infinitely many
fundamental frequencies $\omega _{m}$. However despite its nebulous
character, Bohr used the correspondence principle to very accurately predict
the value of the Rydberg constant $R$ as well as the ``radius'' of a
hydrogen atom.

Bohr's ``old'' quantum theory suffered from a number of defects. In
particular, the increasingly technical quantum conditions seemed unnatural,
and it was difficult to calculate the ``Fourier coefficients $a_{m,n}".$ The
quantity $\left| a_{m,n}\right| ^{2}$ measures the intensity of the
frequencies $\omega _{m,n},$ or at the level of a single atom, to the
probability that a jump from $m$ to $n$ might occur. Just as one cannot ``in
principle'' predict when a radioactive atom might decay, one can cannot say
when an electron will ``jump''. This is a prototypical example of the
probablilisic nature of quantum mechanics.

Heisenberg concluded that the weakness of Bohr's theory rested upon the fact
that it was concerned with predicting the hypothetical singly indexed
energies $E_{n}$ rather than the actually observed doubly indexed
frequencies $\omega _{m,n}.$ As we have seen above, it was this perspective
that led him to consider matrices, . However to carry out his program he had
to incorporate the quantum conditions into his framework.

\section{Phase space and action angle variables}

Quantization is typically applied to algebras of functions. Since the
Hamiltonian approach to mechanics is concerned with an algebra of functions
on a suitable parameter space, it is ideally suited for this process. What
is particularly useful about the Hamiltonian formulation is that each
function determines a one-parameter group of automorphisms, and in
particular, the energy function determines the physical evolution of the
system. Let us summarize this theory as quickly as possible.

Let us first suppose that we are given a parameter space $M=\Bbb{R}^{n}$. We
let $\mathcal{D}(M)$ be the algebra of infinitely differentiable functions
on $M$ and $T(M)=M\times \Bbb{R}^{n}$ be the coresponding tangent space. We
recall that we regard $(x,v)\in TM$ as a ``tangent vector at $x$''$,$ and
that it determines a corresponding directional derivative. Given $x\in M$
and $v=\sum v_{j}e_{j}\in \Bbb{R}^{n},$ we define 
\[
D_{(x,v)}:\mathcal{D}(M)\rightarrow \Bbb{R}:f\mapsto \sum v_{j}\frac{%
\partial f}{\partial x_{j}}(x). 
\]
Since tangent vectors are only used to indicate the directonal derivatives
that they define, we use the notation 
\[
(x,v)=\sum v_{j}\left. \frac{\partial }{\partial x_{j}}\right| _{x} 
\]
A \emph{vector field} is a mapping 
\[
F:M\rightarrow T(M):x\mapsto F(x)\in T_{x}(M), 
\]
and we may write 
\[
F(x)=\sum_{j=1}^{n}F_{j}(x)\frac{\partial }{\partial x_{j}}. 
\]
Given $f\in \mathcal{D}(M),$ we have $D_{F}:x\mapsto D_{F(x)}f$ is again a
smooth function on $M,$ and the mapping 
\[
D=D_{F}:\mathcal{D}(M)\rightarrow \mathcal{D}(M) 
\]
is a derivation of the algebra $\mathcal{D}(M),$ i.e, we have 
\[
D(fg)=D(f)g+fD(g). 
\]
As is well known, all derivations of $\mathcal{D}(M)$ arise in this manner
(see \cite{W}).

A curve 
\[
x:(a,b)\rightarrow M:t\mapsto x(t)=(x_{1}(t),\ldots ,x_{n}(t)) 
\]
is an \emph{integral curve} for a vector field $F$ if for each $t,$ $%
x^{\prime }(t)$ $=$ $F(x(t)).$ Thus $x(t)=(x_{1}(t),\ldots ,x_{n}(t))$ is
just the solution to the system of first order differential equations 
\[
\frac{\mathrm{d}x_{j}(t)}{\mathrm{d}t}=F_{j}(x(t)). 
\]
Under appropriate conditions we may find an integral flow for the vector
field, i.e, a family of mapping $\sigma _{t}:M\rightarrow M$ such that for
each $x\in M$, $x\mapsto \sigma _{t}f$ is an integral curve, and furthermore 
$\sigma _{t+t^{\prime }}=\sigma _{t}\circ \sigma _{t^{\prime }},$ $\sigma
_{0}=I.$ This in turn determines a one-parameter group of algebraic
automorphisms $\alpha _{t}$ of the algebra $\mathcal{D},$ where $\alpha
_{t}f(x)=f(\sigma _{-t}x).$ Using power series one finds a simple
relationship between the derivation $D_{F}$ and the automorphism group $%
\alpha _{t}:$

\begin{eqnarray*}
D_{F}(f) &=&\lim_{h\rightarrow 0}\frac{\alpha _{h}(f)-f}{h} \\
\alpha _{t}(f) &=&e^{tD_{F}}f=\sum \frac{t^{n}}{n!}D_{F}^{n}(f)
\end{eqnarray*}

Turning to physics, let us consider a single oscillating particle with one
degree of freedom. The Newtonian equation of motion is given by $F=ma.$ Let
us assume that the force $F\ $only depends upon the position $q.$ Thus we
are considering the second order equation 
\[
F(q(t))=m\frac{\mathrm{d}^{2}q}{\mathrm{d}t^{2}} 
\]
Since we have restricted to one spatial dimension, $F$ is automatically
conservative, i.e., $F(q)=-V^{\prime }(q)$ for some function $V,$ namely $%
V(q)=-\int F(q)dq$.

We begin by replacing Newton's equation by two first order equations.
Although there are many ways this can be done (e.g. one can let $dq/dt=v,$
and $dv/dt=F/m)$ Hamilton found a particularly elegant way of doing this.
Specifically we use the variables $q$ and $p=mv.$ The corresponding
equations are 
\begin{eqnarray}
\frac{dq}{dt} &=&\frac{\partial H}{\partial p}  \nonumber \\
\frac{dp}{dt} &=&\!\!\!-\frac{\partial H}{\partial q}  \label{Ham}
\end{eqnarray}
where 
\[
H(q,p)=\frac{p^{2}}{2}+V(q). 
\]
We may regard the solutions curves $\gamma (t)=(q(t),p(t))$ as as the
integral curves of the \emph{symplectic gradient }vector field 
\[
\mathrm{sgrad}H=\frac{\partial H}{\partial p}\frac{\partial }{\partial q}-%
\frac{\partial H}{\partial q}\frac{\partial }{\partial p}. 
\]
in the\emph{\ phase space} $M_{2}=\Bbb{R}^{2}$ of variables $(q,p).$ This
quantity is the ``symplectic'' analogue of the usual gradient 
\[
\mathrm{grad}H=\frac{\partial H}{\partial q}\frac{\partial }{\partial q}+%
\frac{\partial H}{\partial p}\frac{\partial }{\partial q}, 
\]
but it is not necessary to go into details.

In fact an arbitrary function $a(q,p)$ on $M_{2}$ determines a vector field 
\[
\mathrm{sgrad}\,a=\frac{\partial a}{\partial p}\frac{\partial }{\partial q}-%
\frac{\partial a}{\partial q}\frac{\partial }{\partial p}, 
\]
and thus corresponding flow 
\[
\sigma _{t}^{a}:M_{2}\rightarrow M_{2}, 
\]
where $\gamma (t)=\sigma _{t}^{a}(x_{0})(q(t),p(t))$ is a solution of the
``Hamiltonian system'' 
\begin{eqnarray*}
\frac{dq}{dt} &=&\frac{\partial a}{\partial p} \\
\frac{dp}{dt} &=&\!\!\!-\frac{\partial a}{\partial q}
\end{eqnarray*}

The \emph{Poisson brackets} of two functions $a$ and $b$ is defined by 
\[
\left\{ a,b\right\} =(\mathrm{sgrad}\,a)(b)=\frac{\partial a}{\partial p}\frac{%
\partial b}{\partial q}-\frac{\partial a}{\partial q}\frac{\partial b}{%
\partial p}. 
\]
In particular, we note that if $\left\{ a,b\right\} =0$, then letting $%
(q(t),p(t))$ be an integral curve of (\ref{Ham}), 
\[
\frac{db}{dt}=\frac{\partial b}{\partial q}\frac{dq}{dt}+\frac{\partial b}{%
\partial p}\frac{dp}{dt}=\frac{\partial b}{\partial q}\frac{\partial a}{%
\partial p}-\frac{\partial b}{\partial p}\frac{\partial a}{\partial q}=0, 
\]
i.e., the function $b$ is constant on the orbits of $a.$ Since $\left\{
a,a\right\} =0,$ we see that $a$ is constant on its own integral curves.

Perhaps the most striking attribute of the phase space parametrization is
that the area $pq$ of a rectangle has the dimensions 
\[
\lbrack \mathrm{momentum}]\times \left[ \text{\textrm{distance}}\right] =%
\mathcal{ML}^{2}\mathcal{\mathcal{\mathcal{T}}}^{-1} 
\]
i.e., \emph{area is an action variable}. This link between the notion of
area (or more precisely the area two-form $\Omega =dp\wedge dq$) and a
physical parameter is one of the most powerful features of the Hamiltonian
theory. We say that a change of variable $Q(q,p),P(q,p)$ is \emph{canonical }
if it preserves the area form, i.e., we have that the Jacobian is 1: 
\[
1=\frac{\partial (Q,P)}{\partial (q,p)}=\frac{\partial Q}{\partial q}\frac{%
\partial P}{\partial p}-\frac{\partial Q}{\partial p}\frac{\partial P}{%
\partial q}. 
\]
If that is the case, then the dynamical system $Q(t)=Q(q(t),p(t))$, $%
P(t)=P(q(t),p(t))$ is also Hamiltonian, i.e., it has the form 
\begin{eqnarray*}
\frac{dQ}{dt} &=&\frac{\partial H}{\partial Q} \\
\frac{dP}{dt} &=&\!\!\!-\frac{\partial H}{\partial Q}
\end{eqnarray*}
where $H(Q,P)=H(q(Q,P),p(Q,P)).$ To see this we note that 
\begin{eqnarray*}
\frac{dQ}{dt} &=&\frac{\partial Q}{\partial q}\frac{dq}{dt}+\frac{\partial Q%
}{\partial p}\frac{dp}{dt} \\
&=&\frac{\partial Q}{\partial q}\frac{\partial H}{\partial p}-\frac{\partial
Q}{\partial p}\frac{\partial H}{\partial q} \\
&=&\frac{\partial Q}{\partial q}\left( \frac{\partial H}{\partial Q}\frac{%
\partial Q}{\partial p}+\frac{\partial H}{\partial P}\frac{\partial P}{%
\partial p}\right) -\frac{\partial Q}{\partial p}\left( \frac{\partial H}{%
\partial Q}\frac{\partial Q}{\partial q}+\frac{\partial H}{\partial P}\frac{%
\partial P}{\partial q}\right) \\
&=&\frac{\partial H}{\partial P}\frac{\partial (Q,P)}{\partial (q,p)}=\frac{%
\partial H}{\partial P},
\end{eqnarray*}
and a similar calculation may be used for the second equation. It is also easy
to see that a canonical change of variables will leave the Poisson brackets
of functions invariant. Given a Hamiltonian co-ordinate system $(Q,P),$ we say that
$Q$ and
$P$ are conjugate variables.

Let us assume that our system is oscillatory, i.e.,
all of the solution curves $(q(t),p(t))$ are closed. We may assume that $%
(q(0),p(0))=(q(T),p(T)),$ where $T$ depends upon the orbit. Our goal is to
find the ``simplest Hamiltonian co-ordinate system'' $(\theta ,I)$ by using
a canonical transformation, with the properties 
\begin{itemize}
\item $H(\theta ,I)=H(I),$ i.e., $H$ doesn't depend upon $\theta ,$ and
\item $\theta $ increases by $2\pi $ on each closed orbit.
\end{itemize}
Given such a system, we have 
\[
\frac{dI}{dt}=-\frac{\partial H}{\partial \theta }=0 
\]
and thus $I$ and $H(I)$ are constant on each orbit $\gamma $. It follows that 
\[
\omega =\frac{d\theta }{\partial t}=\frac{\partial H}{\partial I} 
\]
is also constant on each orbit $\gamma $, i.e., $\omega =\omega (I)=\omega
(\gamma ),$ and $\theta (t)=\omega t+C$ for some constant $C.$
We may assume $C=0,$ and from the second property, $\omega =2\pi /T.$

The canonical transformation from $(q,p)$ to $(\theta ,I)$ transforms the
area $A$ enclosed by an orbit $\gamma (t)=(q(t),p(t))$ to the area $R$ of
the rectangle $0\leq \theta \leq 2\pi ,$ $0\leq I\leq I(\gamma )$. Since the
purported transformation is canonical, we have 
\[
\oint_{\gamma }pdq=A=R=2\pi I(\gamma ), 
\]
where $\gamma $ is the unique integral curve that passes through $(q,p).$
Thus assuming that we can find a canonical transformation with the desired
properties, $I$ is an action variable. For the proof that the transformation
exists (and a formula for $\theta )$ we recommend \cite{PR} or \cite{G}. We
define 
\[
I=\frac{1}{2\pi }\oint_{\gamma }pdq 
\]
to be the \emph{action variable} and $\theta $ the \emph{angle variable. }As
one would expect, $\theta $ is multivalued since it increases by $2\pi $
on each circuit of an orbit.

The action, angle variables enable us to use Fourier series in our
analysis of a periodic motion. Given an arbitrary function $a$ on $M_{2}$
and using the action-angle variables, the function $a(I,\theta )$ will
have period $2%
\pi $ in $\theta .$ It thus has a Fourier series 
\begin{equation}
a(I,\theta )=\sum a(n)e^{in\theta }.  \label{fseries}
\end{equation}
where $a(n)$ is a function of $I.$ Substituting the solution of the
Hamiltonian equations, we obtain $A$ as a function of time: 
\[
a(t)=\sum a(n)e^{in\omega t}
\]
where $a(n)$ is constant on the orbit.

\section{The commutation relation}

We will identify the energy variables $H$ and $E.$ There is a close parallel
between the classical formula 
\[
\omega =\frac{\partial H}{\partial I}=\frac{\partial E}{\partial I} 
\]
and Bohr's difference formula

\[
\omega _{m,m-k}=\frac{E_{m}-E_{m-k}}{\hbar } 
\]
To make this more explicit, let us ``discretize'' the action variable $I$ by
letting $I=m\hbar $ and $\Delta _{k}I=k\hbar .$ Then according to Bohr's
correspondence principle, if $k\ll m,$ 
\[
\frac{\Delta _{k}E}{\Delta _{k}I}=\frac{\Delta _{k}E}{k\hbar }=\frac{%
E_{m}-E_{m-k}}{\hbar }=k^{-1}\omega _{m,m-k}\thicksim k^{-1}k\omega =\frac{%
\partial E}{\partial I}. 
\]
It thus appears that Bohr's correspondence principle is embodied in the fact
that the finite difference with respect to the discrete action variable $I$
approximates the differential quotient with respect to the continuous action
variable $I.$ For this reason it seems justifiable to apply this to
arbitrary quantum variables and their classical analogues. We will use the
symbolism 
\begin{equation}
\frac{\Delta _{k}}{\Delta _{k}I }\leftrightarrow \frac{\partial }{\partial I}
\label{difference}
\end{equation}
(see \cite{W}). The difference operator will be applied to a matrix variable
by the formula 
\[
\left( \Delta _{k}A\right) (m,n)=A(m,n)-A(m-k,n-k). 
\]

In his calculation, Heisenberg concentrated upon the Fourier coefficient
functions $a(\ell)$ of a function $a$ on the phase space $M_{2}$ and the scalar
matrix coefficients $A(m,n)$ of a matrix $A$ in the ``expansions'' 
\begin{eqnarray*}
a &=&\sum_{\ell} a(\ell)e^{i\ell\omega t} \\
A &=&[A(m,n)e^{i\omega _{m,n}t}]
\end{eqnarray*}
If $a$ is the classical function variable ``reduction'' of the matrix
variable $A,$ then for $\ell =m-n\ll m$ the coefficient $A(m,n)$ of $e^{i\omega
_{m,n}t}$ should approximate the coefficient $a(\ell)$ of the overtone $%
(e^{i\omega t})^{\ell}.$ We will write $A\rightsquigarrow a,$ and $%
A(m,n)\rightsquigarrow a(\ell).$ We wish to show that if $A\rightsquigarrow a$
and $B\rightsquigarrow b,$ then $[A,B]\rightsquigarrow \left\{ a,b\right\} .$

If $j,k\ll m$  then  
\begin{eqnarray*}
A(m,m-j) &\rightsquigarrow &a(j)=\frac{1}{i}j^{-1} \frac{\partial a%
}{\partial \theta }(j)\,\, (\mathrm{for}\,\, j\neq0)\\
(\Delta _{k}A)(m,m-j) &\rightsquigarrow &k\hbar \frac{\partial a}{\partial I}%
(j).
\end{eqnarray*}
The equality is seen if one takes the derivative of (\ref{fseries}) with respect to
$\theta$. The second reduction is a formal consequence of (\ref{difference}).

Let us suppose that we are given matrices $A$ and $B\ $and functions $a$ and $b$
with
$ A\rightsquigarrow a$ and $B\rightsquigarrow b$. If
$\ell =m-n\ll m$ 
\begin{eqnarray*}
\lefteqn{(AB-BA)(m,n)} \\
&=&\sum_{j+k=\ell }A(m,m-j)B(m-j,m-j-k) \\
&&\hspace{0.2in}-\sum_{j+k=\ell }B(m,m-k)A(m-k,m-k-j) \\
&=&\sum_{j+k=\ell }[A(m,m-j)-A(m-k,m-j-k)]B(m-j,m-j-k) \\
&&-\sum_{j+k=\ell }A(m-k,m-j-k)[B(m,m-k)-B(m-j,m-j-k)] \\
&=&\sum_{j+k=\ell }(\Delta _{k}A)(m,m-j)B(m-j,m-j-k) \\
&&\hspace{0.5in}-A(m-k,m-k-j)(\Delta _{j}B)(m,m-k) \\
&\rightsquigarrow &\frac{\hbar }{i}\sum_{j+k=\ell }k\frac{\partial a}{
\partial I}(j)\,\,b(k)-a(j)\,\,j\frac{\partial b}{\partial I}(k) \\
&= &\frac{\hbar }{i}\sum_{j+k=\ell,k\neq 0 }k\frac{\partial a}{
\partial I}(j)\,\,k^{-1}\frac{\partial b}{\partial \theta }(k)-\sum_{j+k=\ell,j\neq
0 }j^{-1}\frac{
\partial a}{\partial \theta }(j)\,\,j\frac{\partial b}{\partial I}(k) \\
&=&\frac{\hbar }{i}\left( \frac{\partial a}{\partial I}\frac{\partial b}{%
\partial \theta }-\frac{\partial a}{\partial \theta }\frac{\partial b}{%
\partial I}\right) (\ell ) \\
&=&\frac{\hbar }{i}\left\{ a,b\right\} (\ell ).
\end{eqnarray*}
(see (\ref{convolution}) --- we note that $\frac{\partial b}{\partial \theta
}(0)=\frac{\partial a}{\partial \theta }(0)=0$).

As Heisenberg points out in a footnote, this calculation is
problematical even as a heuristic guide. Although $n-m=\ell =j+k$ is assumed
``relatively small'' with respect to $m$ and $n,$ we are summing  over arbitrary
$j,k$ with $j+k=\ell .$ Heisenberg explains this away by pointing out that if $j$ is
large it will follow that $k$ is large (usually with opposite sign) and vice versa,
and thus all the matrix positions  $(m,m-j),$ $(m-j,m-j-k)$, $(m,m-j),$ and
$(m-k,m-k-j)$ will be distant from the diagonal. He states that the corresponding
matrix elements must be negligible ``since they correspond to high harmonics in the
classical theory''.

We conclude 
\[
\left[ A,B\right] \rightsquigarrow \frac{\hbar }{i}\left\{ a,b\right\} 
\]
Since 
\[
\left\{ p,q\right\} =\frac{\partial p}{\partial p}\frac{\partial q}{\partial %
q}-\frac{\partial p}{\partial q}\frac{\partial q}{\partial p}=1,
\]
if we let $P$ and $Q$ be the quantized momentum matrices, i.e. $%
P\rightsquigarrow p$ and $Q\rightsquigarrow q$, we are led to postulate the
commutation rule 
\[
\left[ P,Q\right] =\frac{\hbar }{i}I.
\]

This relation is the most essential algebraic ingredient of quantum
mechanical computations. The reader may find early instances of these
calculations in \cite{B}.

\end{document}